# Fully Automated Myocardial Strain Estimation from CMR Tagged Images using a Deep Learning Framework in the UK Biobank

Edward Ferdian[a], Avan Suinesiaputra[a], Kenneth Fung,[b], Nay Aung[b], Elena Lukaschuk[c], Ahmet Barutcu[c], Edd Maclean[b], Jose Paiva[b], Stefan K. Piechnik[c], Stefan Neubauer[c], Steffen E Petersen[b], Alistair A. Young[a,d]

[a] Department of Anatomy and Medical Imaging, University of Auckland, Auckland, New Zealand.

[b] William Harvey Research Institute, NIHR Barts Biomedical Research Centre, Queen Mary University of London, Charterhouse Square, London, UK.

[c] Oxford NIHR Biomedical Research Centre, Division of Cardiovascular Medicine, Radcliffe Department of Medicine, University of Oxford, Oxford, UK.

[d] Department of Biomedical Engineering, King's College London, London, UK.

# Abstract

**Purpose:** To demonstrate the feasibility and performance of a fully automated deep learning framework to estimate myocardial strain from short-axis cardiac magnetic resonance tagged images.

**Methods and Materials:** In this retrospective cross-sectional study, 4508 cases from the UK Biobank were split randomly into 3244 training and 812 validation cases, and 452 test cases. Ground truth myocardial landmarks were defined and tracked by manual initialization and correction of deformable image registration using previously validated software with five readers. The fully automatic framework consisted of 1) a convolutional neural network (CNN) for localization, and 2) a combination of a recurrent neural network (RNN) and a CNN to detect and track the myocardial landmarks through the image sequence for each slice. Radial and circumferential strain were then calculated from the motion of the landmarks and averaged on a slice basis.

**Results:** Within the test set, myocardial end-systolic circumferential Green strain errors were -0.001 ± 0.025, -0.001 ± 0.021, and 0.004 ± 0.035 in basal, mid, and apical slices respectively (mean ± std. dev. of differences between predicted and manual strain). The framework reproduced significant reductions in circumferential strain in diabetics, hypertensives, and participants with previous heart attack. Typical processing time was ~260 frames (~13 slices) per second on an NVIDIA Tesla K40 with 12GB RAM, compared with 6-8 minutes per slice for the manual analysis.

**Conclusions:** The fully automated RNNCNN framework for analysis of myocardial strain enabled unbiased strain evaluation in a high-throughput workflow, with similar ability to distinguish impairment due to diabetes, hypertension, and previous heart attack.

# Introduction

Cardiovascular magnetic resonance (CMR) tissue tagging is the non-invasive gold standard for myocardial strain estimation (1-4). Although CMR feature tracking allows calculation of strain from standard steady-state free precession images, features are limited to myocardial edges (5) and structures outside the myocardium (6), whereas CMR tagging enables detection and tracking of features within the myocardium. Displacement encoding with stimulated echoes (DENSE) (7, 8) has the potential to provide higher spatial resolution strain estimates (9), but to date has not been as widely utilized (10). The utility of CMR tagging has been demonstrated in many different patient groups (4). However, there is a lack of robust fully automated analysis tools for the quantification of strain from CMR tagged images, leading to analysis times which are prohibitive in a high-throughput setting, such as studies with many hundreds of cases or high volume clinical centers with >20 cases per week (4, 11, 12).

The most common approaches for strain analysis of CMR tagged images include profile matching and spline fitting (13), deformable contours (14), harmonic phase analysis (15), and sine wave modelling (16). However, these methods require manual initialization and lack robustness. Recently, deep learning methods, in particular convolutional neural networks (CNN), have shown promise for general image processing including automated CMR ventricular function analysis (17-20). However, there have been no reports using neural networks specifically designed for robust analysis of myocardial motion and strain.

In this paper, we developed a fully automated deep learning framework using two neural networks to estimate the left ventricular (LV) circumferential and radial strain on short-axis CMR tagging images. The framework utilized spatial and temporal features to estimate the location and motion of myocardial landmarks, which were placed in consistent anatomical locations regardless of the overlying tag locations. Spatial features were extracted and learned by using a CNN architecture, while the temporal behavior was learned using a recurrent neural network (RNN) (21). This novel spatiotemporal neural network architecture was developed and validated on large scale UK Biobank data (22), and is the first to our knowledge to fully automatically estimate strains from CMR tagging images in a high-throughput setting.

## Methods

### Dataset

This study examined 5,065 UK Biobank participants who underwent CMR examination as part of the pilot phase (April 2014 –August 2015) of the UK Biobank imaging enhancement substudy (22). A previous report described left ventricular shape analysis in this cohort (23). Details of the image acquisition protocol have been described previously (22). This study was approved by the NHS National Research Ethics Service on 17th June 2011 (Ref 11/NW/0382). All participants gave written informed consent. A Siemens Aera 1.5T scanner running Syngo VD13A was used. CMR tagged images comprised gradient recalled echo images acquired in three short-axis slices (basal, mid, and apical) with flip angle 12°, TR/TE 8.2/3.9 msec, field of view 350x241mm, acquisition matrix 256x174, voxel size 1.4x1.4x8.0mm, prospective triggering, tag grid spacing 6mm, temporal resolution 41 msec, ~20 reconstructed frames.

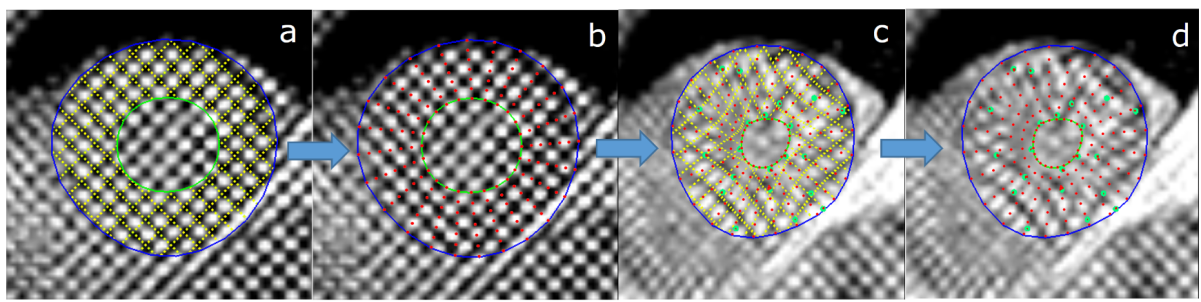

**Figure 1** Manual myocardial landmark generation and tracking. a) Placement of endocardial (green) and epicardial (blue) contours at ED (tag lines shown in yellow); b) myocardial landmark points (red) at ED which were generated automatically; c) tracked tag lines (yellow) at ES, with green points showing manual edits to the displacements; d) final landmarks (red) at ES.

Cases were distributed among five trained readers and the images were analyzed using previously validated software (CIM v6.0, University of Auckland) (9, 24). The majority of the cases (74%) were performed by two career image analysts (JP 41%, EL 33%) whilst the remaining cases were distributed between three cardiologists (AB 12%, KF 9%, EM 5%). The readers had 1-10 years' experience in cardiovascular image analysis. Each reader was trained according to a written standard operating procedure and satisfactorily completed at least 30 training cases prior to contributing towards the ground truth. Figure 1 illustrates the step-by-step process of generating the ground truth landmarks. The software identified and tracked 168 landmarks inside the myocardium at standard anatomical locations, beginning from the mid-point of the septum (half way between the anterior and posterior RV insertion points). The landmarks were equally spaced within the myocardium, with 7 points in the radial (transmural) direction and 24 points in each circumference. The software used a deformable registration algorithm which attempted to track the tags by minimizing the sum of squared differences between consecutive frames (9, 24). The readers manually corrected the tracking to match the motion of the image tags in several key frames: end-diastole (ED, the first frame after detection of the R wave), end-systole, (ES, the frame of maximum contraction), after rapid filling, and at the end of the cycle. The software interpolated these corrections to the intermediate frames. Basal slices were not analyzed if the total circumference of the myocardium affected by presence of left ventricular outflow tract was ≥ 25%; apical slices were not analyzed if there was no evidence of cavity at ES. Cases were also excluded if the tagging image quality was deemed unacceptable by the readers. This resulted in 4,508 CMR tagging cases (12,409 slices), each with 168 landmark points tracked in each frame, available as ground truth for our neural networks. Participants with high blood pressure, diabetes and previous heart attack were identified from the questionnaire data as self-reported existing conditions, or conditions diagnosed by a physician, or taking medications for these conditions. Participant characteristics are shown in Table 1.

## Framework Overview

The deep learning framework consisted of the following steps: 1) find the region of interest (ROI) containing the left ventricular myocardium, 2) crop and resample the image ROI for every frame, 3) detect and track myocardial landmarks across all the frames, and 4) calculate strains based on the motion of the landmarks (Figure 2).

Prior to input into the process, the images were zero-padded to 256x256 matrix size, and each cine (sequence of frames in a slice) was fixed to 20 frames in length. Most slices (n=12,355) already had 20 frames; slices with less than 20 frames (n=13) were padded with empty frames, and slices with greater than 20 frames (n=40) were truncated by taking the first 20 frames.

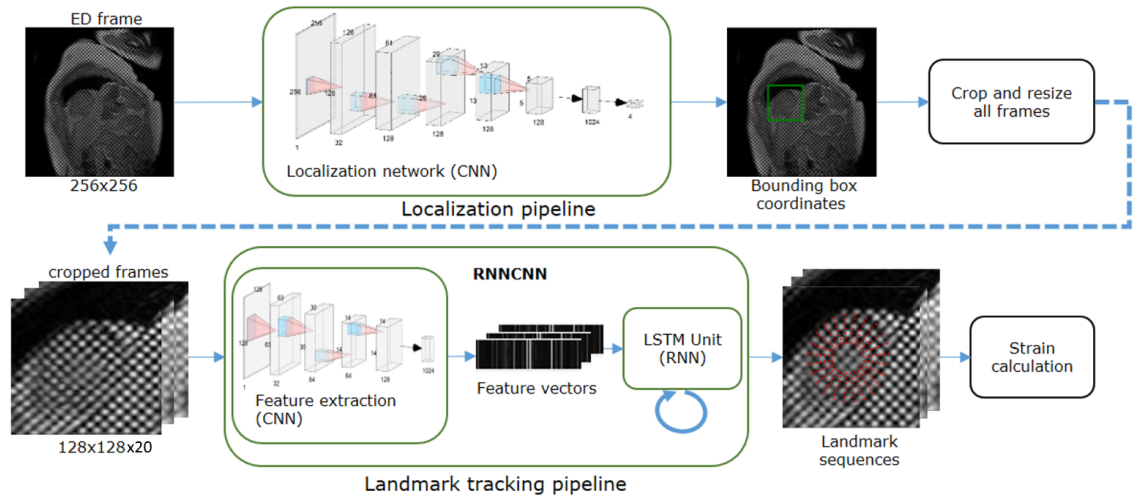

**Figure 2** Overview of the machine learning framework for automatic myocardial strain estimation from CMR tagging.

The neural networks were developed using Tensorflow 1.5.0 (25) and Python, and trained on an NVIDIA Tesla K40 with 12GB RAM. The final output of the framework was radial and circumferential strain, which were calculated from the displacement of landmark points for every time frame using the Green (Lagrangian) strain formula; compatible with finite strain tensors (9, 26):

$$\varepsilon(t) = \frac{1}{2}\left(\frac{L_t^2 - L_0^2}{L_0^2}\right) \quad (1)$$

where $L_t$ represents the segment length at any frame $t$, $L_0$ represents the initial length.

During training, the data were randomly divided by case into 90% training and validation set (n=4056 cases), and 10% test set (n=452 cases). The first set were further partitioned into 80% training (n=3244 cases) and 20% validation (n= 812 cases), used for checking overfitting and convergence and tuning model parameters.

## ROI Localization

The localization network was designed to detect the ROI enclosing the LV myocardium in the ED frame. The output of this network was a rectangular bounding box defined by the extent of the myocardium, with a 60% increase to ensure enough spatial information was included from outside the myocardium. The network configuration is shown in Supplementary Figure 1. Each convolution layer was followed by batch normalization (27). Rectified linear unit (ReLU) was used as activation function on every layer (28), except for the output layer which was a regression layer. A dropout layer (29) with 20% dropout probability was utilized after the first fully connected layer.

The network was optimized using the mean squared error (MSE) between the prediction and ground-truth bounding-box corners as the loss function. The Adam optimizer with a learning rate of $10^{-3}$ was used; learning rate was reduced by a factor of $\sqrt{2}$ for every $5^{th}$ epoch after the $10^{th}$ epoch. Accuracy was calculated using the intersection over union (IoU) metric, defined to be the area of overlap of the predicted and ground truth bounding box, divided by the union of the areas of the predicted and ground truth boxes.

After the ED frame bounding box was obtained, all the images in the cine were cropped using the same bounding box. Since the heart at ED is fully expanded before contraction, myocardium at the following frames after ED has smaller area; hence ensuring the bounding box covers the myocardium in all frames. Subsequently all the cropped images were resampled to 128x128 pixels using bicubic interpolation to be fed into the landmark tracking network.

## Landmark Tracking

The landmark tracking network (RNNCNN) was constructed from two components, a CNN component designed to extract the spatial features, and an RNN component designed to incorporate the temporal relationship between frames. The input data for this network consisted of 20 frames of size 128x128 pixels, taken from the output of the localization pipeline. The RNNCNN was trained end-to-end as a single network. Training time took approximately 10 hours.

A summary of the RNNCNN architecture is shown in Supplementary Figure 2. Leaky ReLU (30) activation function (with α=0.1) was used in the shared-weight CNN component. The CNN component took one frame at a time and output a 1024 length feature vector per frame. The dynamic RNN (max 20 frames) used a Long Short-Term Memory (LSTM) unit (31) with 1024 nodes. ReLU was used as activation function in the RNN component. The final output layer was a regression layer, resulting in 168 landmark coordinates for 20 time frames.

The RNNCNN was optimized using a composite loss function that simultaneously minimized position error, radial and midwall circumferential strain errors in each frame, defined on a slice-by-slice basis as follows (Figure 3):

$$loss_t = MSE_t + \omega * \text{error}(\varepsilon_{R(t)}) + \omega * \text{error}(\varepsilon_{c(t)})$$

$$MSE = \frac{1}{n} \sum_{i=1}^{n} (x'_i - x_i)^2 + (y'_i - y_i)^2$$

$$\text{error}(\varepsilon_R) = |\varepsilon'_R - \varepsilon_R| \qquad \text{error}(\varepsilon_c) = |\varepsilon'_c - \varepsilon_c|$$

$$\varepsilon_{R(t)} = \frac{1}{24}\frac{1}{2}\sum_{k=1}^{24}\left(\frac{\Delta R_{kt}^2}{\Delta R_{k1}^2} - 1\right) \qquad \varepsilon_{c(t)} = \frac{1}{24}\frac{1}{2}\sum_{k=1}^{24}\left(\frac{\Delta C_{kt}^2}{\Delta C_{k1}^2} - 1\right)$$

(2)

where $MSE_t$ is the mean squared error between predicted ($x_i'$, $y_i'$) and ground truth ($x_i$, $y_i$) landmark positions at frame t (*n* is 168, the number of landmarks in one frame). $\Delta R_{kt}$ is the distance between the epicardial and the endocardial landmarks along each radial line *k* in frame *t*, with *t*=1 being used as the reference frame. $\Delta C_{kt}$ is the distance between two consecutive landmarks in the midwall circumference $k$ at frame $t$ (see Figure 3). The strain errors were given weight ω to adjust for their relative scale compared to the displacement errors. The radial and circumferential strains ($ε_R$ and $ε_C$ respectively) were calculated using the Green strain formula (eqn 1). The strains were averaged over the slice before computing the error (eqn 2).

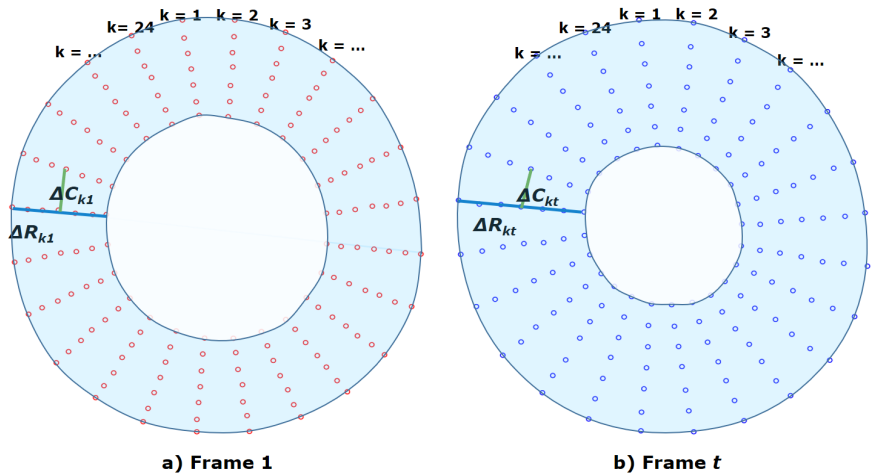

a) Frame 1          b) Frame $t$

**Figure 3** Measurements of radial and circumferential inter-landmark distances at a) frame 1 (ED, assumed as the reference frame) and b) frame $t$. Both images depict the 7 circumferential rings of landmarks. Subendocardial, midwall, and subepicardial circumferential strain was calculated from the $2^{nd}$, $4^{th}$ and $6^{th}$ rings from the center respectively. $ΔR_{kt}$ shows the distance for radial line $k$ in frame $t$, while $ΔC_{kt}$ shows the distance for circumferential line $k$ in frame $t$.

Based on the loss function, the network was effectively optimized using position (MSE) and strain (radial and circumferential) constraints. The Adam optimizer with learning rate $10^{-4}$ was used; learning rate was reduced by a factor of √2 for every $10^{th}$ epoch. Overall accuracy in the test set was calculated based on 1) slice-based strain errors at ES between the predicted strain and ground truth, and 2) root mean squared position error of all landmarks within a slice at ED and ES.

## Statistics

All statistical analyses were performed using SciPy Statistics (32), an open-source Python library for statistical functions. Continuous variables were expressed as mean ± standard deviation, with errors expressed as mean difference ± standard deviation of the differences, computed over slices. We use the terms *bias* to denote the mean difference and *precision* to denote the standard deviation of the differences. These were calculated across basal, mid and apical slices separately, to give results for each location. Bland-Altman analysis was used to quantify agreement by plotting difference against the mean of both measurements. Differences between automated and manual results, and inter-observer differences, were assessed using Student's t test. Bonferroni correction was used with 15 tests (Table 2), giving p<0.0033 as significant. Manual inter-observer errors were obtained by comparing the landmark coordinates and strain differences (mean difference ± standard deviation of the differences, calculated over slices) between two observers on 40 cases. Differences in midventricular circumferential strain due to disease processes (diabetes, high blood pressure, and previous heart attack) were tested using Welch's unequal variances t-test.

# Results

## ROI Localization

The performance of the localization network was evaluated visually and quantitatively on the test dataset (1245 slices). For visual evaluation, we reviewed the cases with the worst IoU and checked whether the cropped ROI was acceptable (i.e. contained the LV myocardium). None of the test cases were deemed to have an unacceptable ROI. The worst case had an IoU=53% and the average IoU for the test set was 90.4 ± 5.4%. Figure 4 shows the worst and typical example results and the accuracy distribution of the localization network.

## Landmark Tracking

The root mean squared position error of the 168 landmark coordinates on the test dataset was 4.1 ± 2.0 mm at ED and 3.8 ± 1.7 mm at ES (mean over slices ± std dev over slices). In comparison, the inter-observer position error on the 40 cases was 2.1 ± 1.8 mm and 2.0 ± 1.6 mm at ED and ES, respectively (mean ± std dev). Root mean squared position error was mainly due to variation in the placement of landmarks in the circumferential direction on the

ED frame, corresponding to the localization of the mid-point of the septum from the RV insertion points in the ground truth.

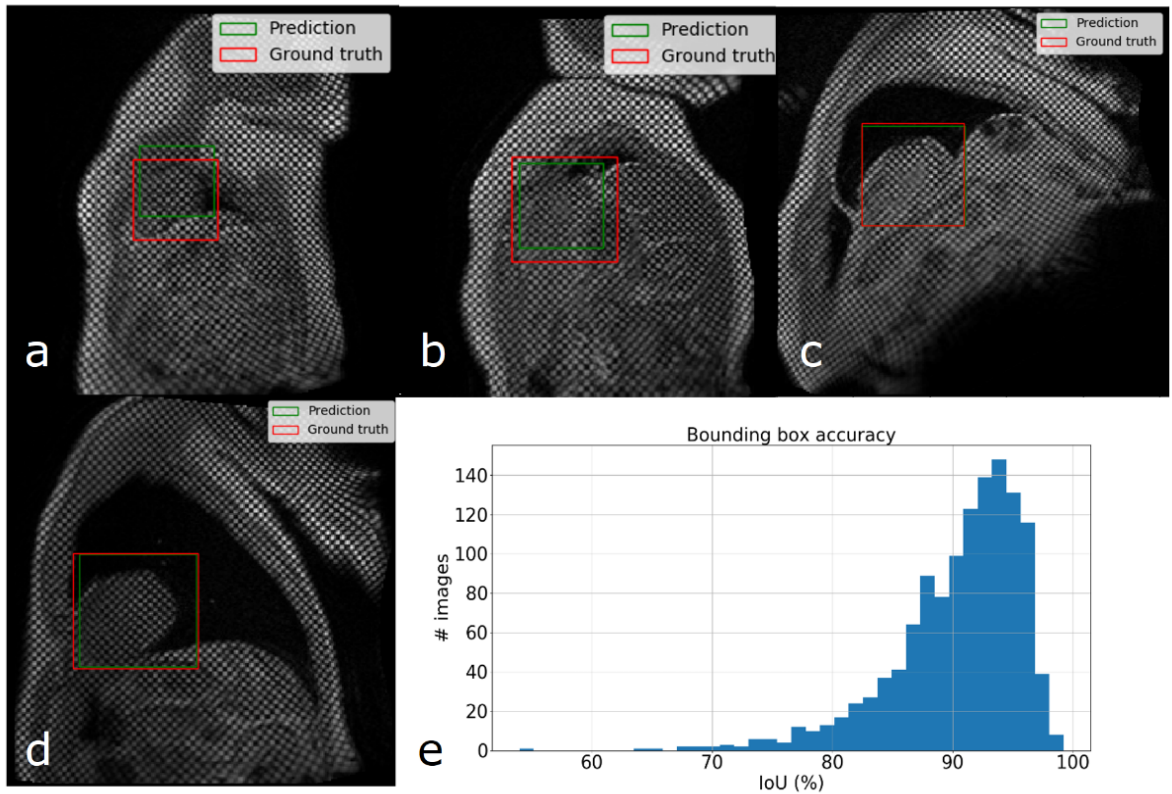

**Figure 4** Example results from the localisation network. a) IoU=53% and b) IoU=63% are the worst prediction results within the test set; c) and d) show typical prediction results; e) histogram showing the distribution of accuracy (IoU) throughout the test set.

Visual checks were performed on the worst cases (IoU < 70%, position error > 4mm, $|error(\varepsilon_R)| > 0.2$, or $|error(\varepsilon_c)| > 0.05$) as well as randomly selected cases to verify the landmark positions were acceptable (i.e. were located in the myocardium). Only 1 out of 1245 slices (0.08%) had grossly misplaced landmarks, due to the inaccurate prediction of the ROI (IoU=53%) which caused the ground truth landmarks (16 out of 168 in the ED frame) located outside the predicted ROI. Table 2 shows the predicted and ground truth strain values, with the mean and standard deviations of the differences between predicted and ground truth strains. In addition to the average strain over the whole myocardium, circumferential strain was also calculated for subepicardial, midwall, and subendocardial regions, using the 2$^{nd}$, 4$^{th}$, and 6$^{th}$ circumferential ring of landmarks respectively, as shown in Figure 3. All strain biases (mean difference) were small and not significantly different from zero, except for basal radial

strain and basal and midventricle midwall circumferential strain (likely due to the relatively large number of image slices). The prediction precision (standard deviation of the differences) was worse than the inter-observer precision, indicating the observers were more precise than the network. As a further comparison, the absolute errors for the network, as well as the absolute errors between the two observers, are shown in Table 2. The inter-observer absolute errors were similar to the network absolute error. Circumferential strains were highest for subendocardial, and lowest for subepicardial regions, in agreement with previous studies (33). Subendocardial estimates showed better precision, with radial strains having the worst precision, in agreement with previous studies (9).

Indicative inter-observer strain errors for two readers (n=40) are also shown in Table 2 for comparison (similar inter-observer differences were observed between the other readers). Small but statistically significant differences in strain were mainly due to differences in contour placement. Similar inter-observer differences using the same software in patients was found previously (9). The automatically predicted strain showed comparable errors to the manual inter-observer strain errors, with worse precision but improved bias over a larger number of cases. As a further comparison, Table 2 also shows errors arising from the deformable registration method (using CIM 6.0) without manual correction on the 40 cases used to calculate the inter-observer variabilities, using the average observer strain as the ground truth. These errors were larger than the automated RNN-CNN prediction.

Figure 5 shows Bland-Altman plots comparing the difference between the predicted and the ground truth ES strains obtained from the landmarks. These confirm that the average of the differences for $\varepsilon_R$ and $\varepsilon_C$ were approximately zero. Most of the cases can be seen to lie within the 95% limits of agreement (bias ± 1.96*precision). Some outliers can be seen indicating a few cases with large error. The limits of agreement for $\varepsilon_R$ were the widest, in agreement with previous studies which showed reduced accuracy for radial strain using CMR tagging (9). We observed smallest limits of agreement for the $\varepsilon_C$ on the middle slice, and largest for the apical slice.

An example of the resulting landmark detection and tracking at ED and ES, including the strain estimations for all time frames, is shown in Figure 6. Tracking error tended to increase toward the end of the cine sequence when tag fading typically occurs during diastole. The fully automated framework could process images at approximately 260 frames (~13 slices) per second in an NVIDIA Tesla K40 with 12GB RAM. In comparison the manual analysis typically required 6-8 minutes per slice, giving an improvement approximately 5000x.

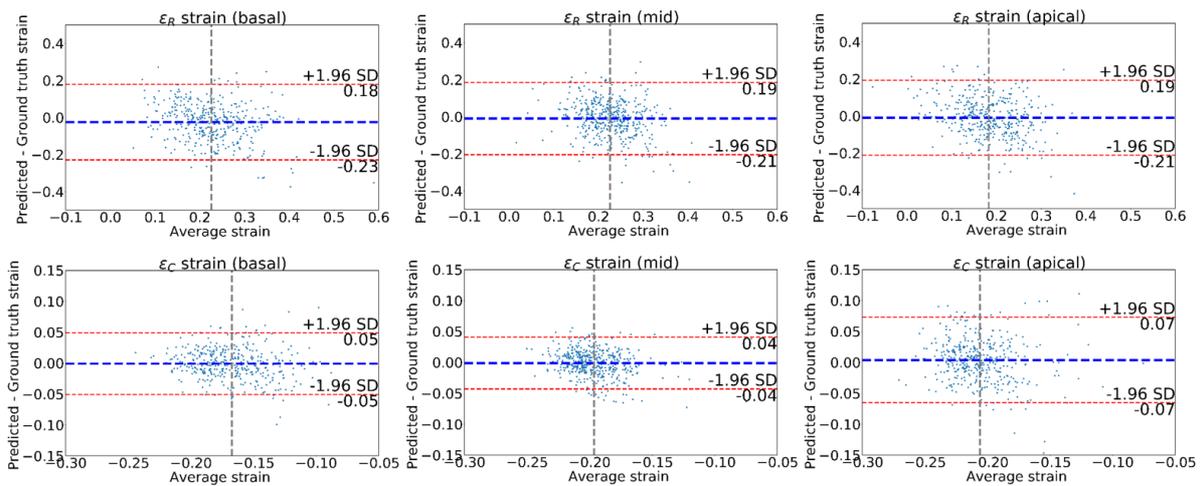

**Figure 5** Bland-Altman plots of the strains obtained for the left ventricle at end systole frame. The strain values obtained from the predicted landmarks were compared to the strains from the ground truth landmark. First row shows the radial strains for 3 different short-axis slices; the second row shows the average circumferential strains. The dashed blue line denotes the mean difference; the dashed red lines denote the 95% limits of agreement (mean $\pm$ 1.96 * standard deviation)

Table 3 shows differences in mid-ventricular circumferential strain for diabetes, high blood pressure and previous heart attack, as self-reported by the UK Biobank participants. In each comparison the manual analysis found statistically significant impairment between disease and reference (cases without hypertension, diabetes or previous heart attack), which were reproduced with similar confidence intervals in each disease by the fully automatic method. LV mass and volume from MRI (34) are shown for comparison (indexed by body surface area). For high blood pressure, there were 818 cases in the training set, 195 cases in the validation set and 108 cases in the test set. For diabetes, this was 135, 30, 16 respectively and for previous heart attack 64, 13, 11 respectively.

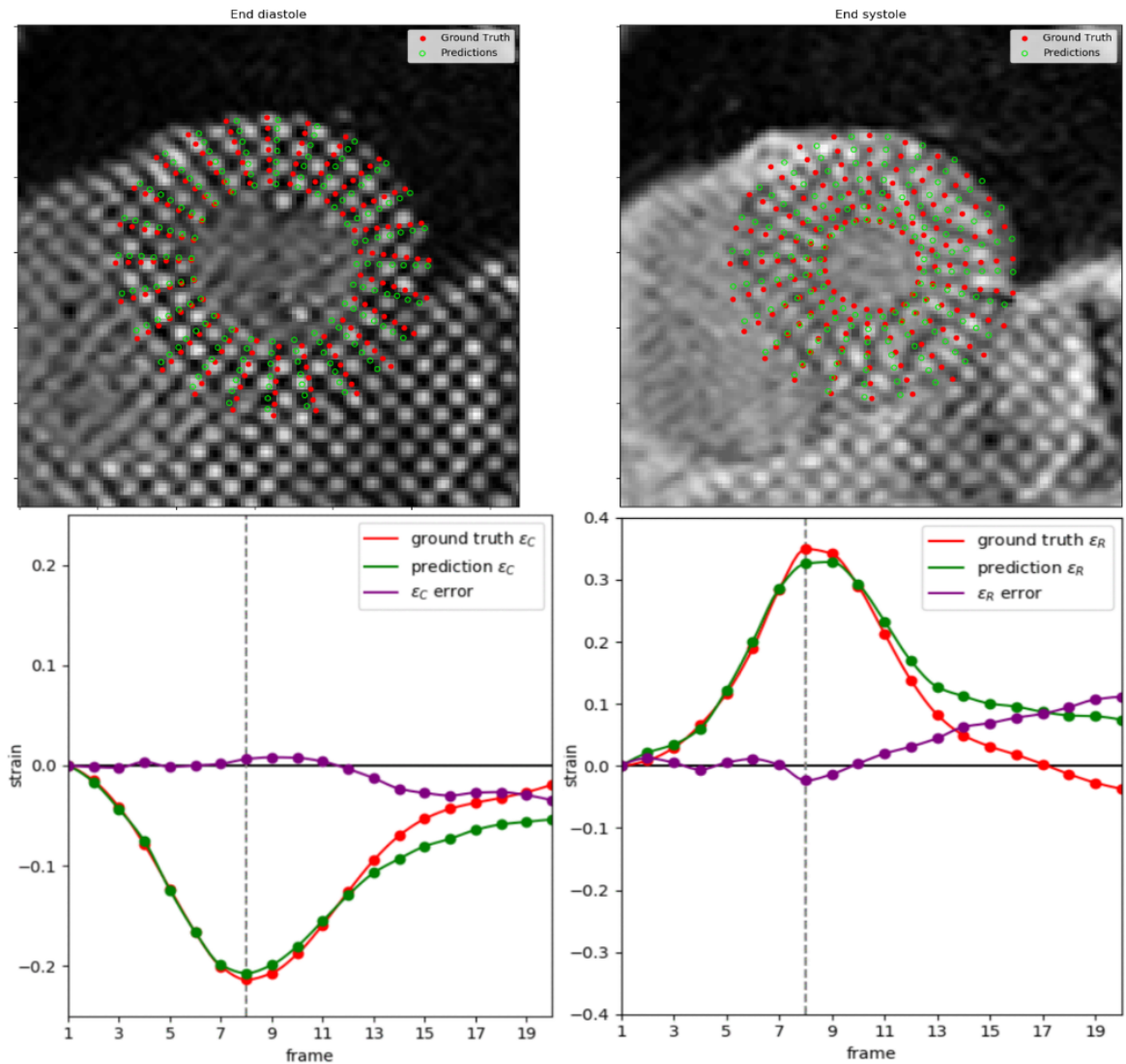

**Figure 6** Example of ground truth compared to estimated landmarks during ED and ES (top row) and strain calculation for the whole cine; circumferential strain (bottom left) and radial strain (bottom right). Dashed vertical line marks the ES frame.

## Discussion

Strain estimation from CMR tagging images is a challenging problem. We have designed a fully automated framework to calculate strains from CMR tagging images, while also providing anatomical landmark points. To our knowledge, our study is the first to provide fully automated analysis in a high-throughput setting. The method is feasible for direct application to the 100,000 participants of the UK Biobank imaging substudy, as these cases become available. The method was able to learn the previously validated deformable registration method, as well as the manual correction of tracking errors. We are not aware of other fully automated methods that do not require some manual intervention for tagged CMR images in practice. Our results suggest that the proposed framework can instantly (in real time) produce unbiased estimates of regional myocardial strains with reasonable precision, which reproduce differences due to disease processes. In a high-throughput clinical setting, this method can be employed as a robust first pass evaluation.

The performance of the fully automated framework is comparable to previous studies comparing CMR tagging with DENSE or feature tracking (9, 11, 35). In particular, a previous validation study using the same manually corrected tagging analysis procedure in patients (9), found inter-observer circumferential strain errors of -0.006 ± 0.034 for tagging, and similar errors between DENSE and tagging (9). Radial strains are known to be underestimated with tagging compared with DENSE and feature tracking (9), and have worse precision due to the large tag spacing relative to the distance between the endocardium and epicardium (3). In our study, both circumferential and radial strains at ES showed minimal bias. Although the basal $\varepsilon_R$, basal midwall $\varepsilon_C$, and the midventricular midwall $\varepsilon_C$ were statistically different ($p<0.0033$), the magnitude of the bias (-0.025, -0.004, and 0.003 respectively) are unlikely to be clinically significant. Since the network saw cases from all readers in the training set, it could learn an average of all readers and avoid the particular bias commonly associated with individual human readers.

CMR tagging has also been incorporated into several large cohort studies, including the Multi-Ethnic Study of Atherosclerosis (36), and the UK Biobank CMR imaging extension (22). In particular, CMR tagging was included in the UK Biobank CMR imaging protocol in order to provide accurate myocardial strain estimation for the analysis of developing disease (22). Fully automatic strain analysis would therefore improve the utility of CMR tagging in large cohort studies. This method can be applied for the automated analysis of the remaining UK Biobank CMR cohorts, which is estimated to be ~100,000 participants by 2023 and 10,000 with repeat imaging two to three years after the initial imaging visit.

The root mean squared position error were relatively high due to the difficulty in locating the precise mid-point of the septum. This was also seen in the inter-observer position error. The error in position does not propagate to errors in strain, which depends only on the relative motions between ED and ES. However, it can be seen that strain errors are increased in the apical slice, due primarily to thin and obliquely oriented myocardial walls, partial voluming, and large motions.

**Experiments and hyper-parameter tuning**

For the localization CNN, lower adjustment fraction ($\lambda<0.3$) gave more precise estimate of the LV ROI, however it left the tracking network with reduced spatial context, which led to reduced accuracy. In future work, further augmentation including arbitrary rotations and translation might further improve the accuracy. However, we found 90% IoU accuracy to be acceptable and no cases were found to fail outright (i.e. not include the LV).

For the landmark tracking network, we also experimented with a separately trained landmark detection CNN and subsequent landmark tracking RNN. The separately trained networks were inferior in performance since the CNN can only process each frame independently, therefore did not guarantee motion coherence from frame to frame. The end-to-end RNNCNN architecture was more difficult to optimize and sensitive to changes. Tweaking the network required a balance between the CNN and RNN components. Weights for the strain errors ($\omega=1, 5, 10$), batch size (20, 25, and 30), CNN activation function (ReLU, leaky ReLU), LSTM nodes (400, 512, 600, 800, 1024, and 2048), and additional dense layers before and after the LSTM unit, resulted in reduced performance during training. In the combined RNNCNN network, we found that leaky ReLU (37) was a key hyperparameter in the CNN component, which allows negative values to be updated and prevents missing spatial information that might be useful for the RNN component. The overall lean RNNCNN architecture was inspired by MariFlow (38).

**Limitations and Future Work**

The neural networks for this study were trained with a single dataset, derived from the UK Biobank, which is homogenous in imaging protocol and consisted of mainly healthy subjects (34). Additional augmentations to the dataset are needed to adapt the neural network for a different dataset, such as pathologies not seen in UK Biobank or data from different imaging protocols. Although the bias was excellent, more work is needed to improve the precision to match that of manual analysis. Table 2 shows that the network precision is approximately twice that of the inter-observer error, which is adequate for large scale studies like UK Biobank but not for identifying subtle changes in individual patients. Another limitation is the number of frames which is fixed (t=20) due to the nature of the current Tensorflow implementation. Future work will explore calculation of segmental strains by assigning a segment label (according to American Heart Association (39)) to every landmark. Additionally, other variants of RNN, such as Bidirectional RNN (40) and Convolutional LSTM (41) are possible candidates to improve the network by allowing backward temporal relationship and preserve spatial information, respectively. It would also be very useful to have an automatic evaluation of tag image quality, particular in the context of tag fading in order to determine when tags are not analyzable in part of the cardiac cycle.

**Code and Data Availability**

The code will be made available at https://github.com/EdwardFerdian/mri-tagging-strain. In addition, the raw data, the derived data, the analysis and results of this study will be available from the UK Biobank central repository (application number 2964). Researchers can request access to these data through the UK Biobank application procedure (http://www.ukbiobank.ac.uk/register-apply/)

**Conclusions**

In this study, we have introduced a fully automated framework to estimate radial and regional circumferential strains from CMR tagging images using a deep learning framework. The framework could detect and track 168 landmarks over many frames by utilizing spatial and temporal features. The method resulted in unbiased estimates of reasonable precision, suitable for a robust evaluation in a high-throughput setting in which manual initialization or interaction is not possible. The method reproduced significant reductions in strain due to diabetes, hypertension and previous heart attack.

***Acknowledgements:*** *This research has been conducted using the UK Biobank Resource under Application 2964. The authors wish to thank all UK Biobank participants and staff.*

***Funding:*** *This research has been conducted using the UK Biobank Resource under application 2964. Funding was provided by the British Heart Foundation (PG/14/89/31194), and by the National Institutes of Health (USA) 1R01HL121754. SN acknowledges the National Institute for Health Research (NIHR) Oxford Biomedical Research Centre based at The Oxford University Hospitals Trust at the University of Oxford, and the British Heart Foundation Centre of Research Excellence. AY acknowledges Health Research Council of New Zealand grant 17/234. AL and SEP acknowledge support from the NIHR Barts Biomedical Research Centre and from the "SmartHeart" EPSRC program grant (EP/P001009/1). NA is supported by a Wellcome Trust Research Training Fellowship (203553/Z/Z). This project was enabled through access to the MRC eMedLab Medical Bioinformatics infrastructure, supported by the Medical Research Council (grant number MR/L016311/1). KF is supported by The Medical College of Saint Bartholomew's Hospital Trust, an independent registered charity that promotes and advances medical and dental education and research at Barts and The London School of Medicine and Dentistry. The UK Biobank was established by the Wellcome Trust medical charity, Medical Research Council, Department of Health, Scottish Government and the Northwest Regional Development Agency. It has also received funding from the Welsh Assembly Government and the British Heart Foundation.*

***Disclosures:*** *SEP provides consultancy to Circle Cardiovascular Imaging Inc, Calgary, Canada. AY has received consulting fees from Siemens Healthineers. EF received partial stipend support from Siemens Healthineers. Other authors have no conflicts of interest to declare.*

# Tables

**Table 1.** Participant characteristics (n=4508). Values are given as mean ± standard deviation for continuous variables, and count (%) for categorical variables.

| | |
|---|---:|
| **Age (years)** | 62 ± 8 |
| **Sex (male)** | 2100 (47%) |
| **Height (cm)** | 170 ± 9 |
| **Weight (kg)** | 75 ± 15 |
| **Body surface area (m$^2$)** | 1.85 ± 0.20 |
| **Systolic blood pressure (mmHg)** | 138 ± 19 |
| **Diastolic blood pressure (mmHg)** | 79 ± 11 |
| **Heart Rate (bpm)** | 68 ± 11 |
| **High blood pressure** | 1130 (25%) |
| **Diabetes** | 182 (4%) |
| **Previous heart attack** | 89 (2%) |

**Table 2** End-systolic (ES) circumferential and radial strains (mean ± standard deviation) in short-axis slices, split between basal, mid, and apical slices. Asterisk (*) indicates statistically significant differences between two measurements (p<0.0033), for *predicted strain error* = predicted–ground truth slice strain, and *manual inter-observer error* = observer1 – observer2 slice strain. For both prediction and inter-observer errors, results are presented at mean difference ± standard deviation of the differences. For absolute errors, the absolute value of the difference is used. Number of slices is shown for each region, except for manual inter-observer error and uncorrected deformable registration method (basal n=35, mid n=40, apical n=34).

| Type/Region | Ground truth | Prediction | Error | | Absolute Error | | |
| --- | --- | --- | --- | --- | --- | --- | --- |
| | | | Prediction | Inter-observer | Prediction | Inter-observer | Uncorrected deformable registration |
| *Basal (n=386)* | | | | | | | |
| $\varepsilon_C$ | -0.167 ± 0.032 | -0.168 ± 0.029 | -0.001 ± 0.025 | -0.018 ± 0.009* | 0.019 ± 0.017 | 0.019 ± 0.009 | 0.043 ± 0.018 |
| Subendo | -0.220 ± 0.045 | -0.220 ± 0.040 | 0.000 ± 0.035 | -0.028 ± 0.012* | 0.027 ± 0.022 | 0.028 ± 0.012 | 0.068 ± 0.025 |
| Midwall | -0.162 ± 0.032 | -0.166 ± 0.027 | -0.004 ± 0.025* | -0.016 ± 0.009* | 0.019 ± 0.016 | 0.016 ± 0.008 | 0.039 ± 0.017 |
| Subepi | -0.114 ± 0.024 | -0.115 ± 0.020 | -0.001 ± 0.021 | -0.008 ± 0.011* | 0.016 ± 0.013 | 0.011 ± 0.008 | 0.019 ± 0.013 |
| $\varepsilon_R$ | 0.238 ± 0.099 | 0.213 ± 0.076 | -0.025 ± 0.104* | -0.023 ± 0.065 | 0.082 ± 0.069 | 0.054 ± 0.043 | 0.071 ± 0.054 |
| *Mid (n=451)* | | | | | | | |
| $\varepsilon_C$ | -0.196 ± 0.025 | -0.197 ± 0.021 | -0.001 ± 0.021 | -0.011 ± 0.016* | 0.016 ± 0.014 | 0.016 ± 0.011 | 0.049 ± 0.019 |
| Subendo | -0.256 ± 0.033 | -0.258 ± 0.028 | -0.002 ± 0.030 | -0.020 ± 0.020* | 0.023 ± 0.020 | 0.025 ± 0.015 | 0.072 ± 0.022 |
| Midwall | -0.191 ± 0.025 | -0.194 ± 0.020 | -0.003 ± 0.021* | -0.008 ± 0.016 | 0.016 ± 0.014 | 0.014 ± 0.011 | 0.045 ± 0.019 |
| Subepi | -0.135 ± 0.022 | -0.135 ± 0.017 | 0.000 ± 0.018 | -0.001 ± 0.015 | 0.014 ± 0.012 | 0.011 ± 0.009 | 0.029 ± 0.016 |
| $\varepsilon_R$ | 0.231 ± 0.085 | 0.221 ± 0.064 | -0.010 ± 0.100 | -0.015 ± 0.053 | 0.076 ± 0.065 | 0.046 ± 0.030 | 0.083 ± 0.058 |
| *Apical (n=408)* | | | | | | | |
| $\varepsilon_C$ | -0.208 ± 0.032 | -0.204 ± 0.029 | 0.004 ± 0.035 | -0.013 ± 0.016* | 0.027 ± 0.023 | 0.015 ± 0.013 | 0.073 ± 0.027 |
| Subendo | -0.273 ± 0.043 | -0.268 ± 0.039 | 0.005 ± 0.049 | -0.026 ± 0.019* | 0.037 ± 0.033 | 0.027 ± 0.018 | 0.107 ± 0.034 |
| Midwall | -0.202 ± 0.032 | -0.200 ± 0.027 | 0.001 ± 0.034 | -0.009 ± 0.016* | 0.026 ± 0.022 | 0.014 ± 0.012 | 0.066 ± 0.026 |
| Subepi | -0.143 ± 0.028 | -0.139 ± 0.021 | 0.003 ± 0.028 | 0.000 ± 0.018 | 0.022 ± 0.018 | 0.014 ± 0.012 | 0.042 ± 0.024 |
| $\varepsilon_R$ | 0.187 ± 0.096 | 0.178 ± 0.071 | -0.009 ± 0.103 | -0.002 ± 0.080 | 0.081 ± 0.064 | 0.062 ± 0.051 | 0.128 ± 0.082 |

**Table 3** Reduction in midventricular slice circumferential strain associated with diabetes, hypertension or previous heart attack, in 1245 midventricular slices from the test and validation cases. CI: 95% confidence interval for the differences between disease and reference (Welch two-sample t-test with unequal variances). Reference: cases without hypertension, diabetes or previous heart attack. LV volumes and mass also given for comparison. EDVi: end-diastolic volume index; ESVi: end-systolic volume index; LVMi: LV mass index; EF: ejection fraction. Volumes and mass are indexed to body surface area. *P<0.05 compared with Reference.

|  | Diabetes (n=46) | Hypertension (n=303) | Heart Attack (n=24) | Reference (n=888) |
|---|---|---|---|---|
| $\varepsilon_C$ (Manual) | -0.188 ± 0.030* | -0.192 ± 0.025* | -0.175 ± 0.040* | -0.199 ± 0.023 |
| CI | (0.003, 0.020) | (0.004, 0.010) | (0.008, 0.042) | |
| $\varepsilon_C$ (Pred.) | -0.186 ± 0.020* | -0.194 ± 0.022* | -0.176 ± 0.033* | -0.199 ± 0.022 |
| CI | (0.006, 0.019) | (0.002, 0.008) | (0.009, 0.037) | |
| EDVi (ml/m$^2$) | 75 ± 12 | 78 ± 14 | 88 ± 14* | 78 ± 15 |
| ESVi (ml/m$^2$) | 33 ± 11 | 32 ± 9 | 40 ± 11* | 32 ± 9 |
| LVMi (g/m$^2$) | 50 ± 10* | 51 ± 10* | 49 ± 9 | 46 ± 9 |
| EF (%) | 58 ± 10 | 59 ± 7 | 53 ± 8* | 60 ± 6 |

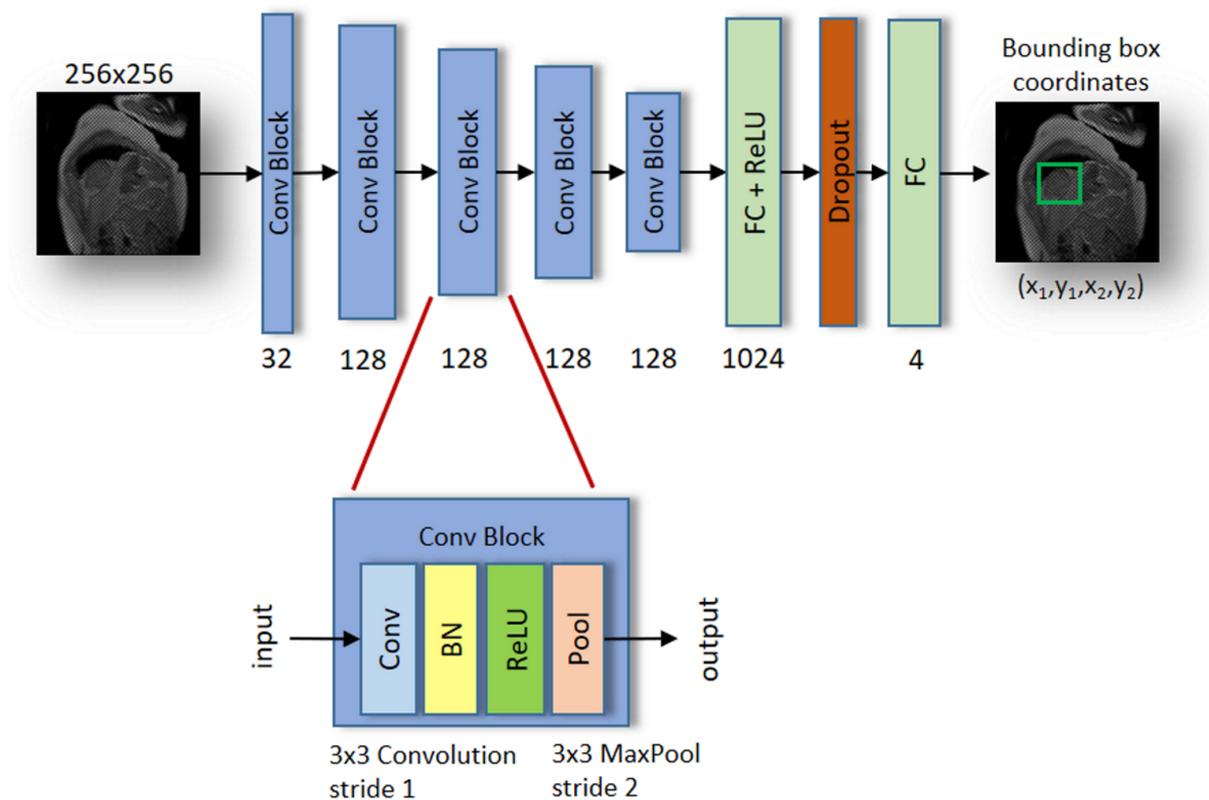

**Supplementary Figure 1** Localization network architecture (top). Convolution block (bottom). Conv = convolution, BN = batch normalization, ReLU = rectified linear unit, Pool = pooling layer, FC = fully connected layer.

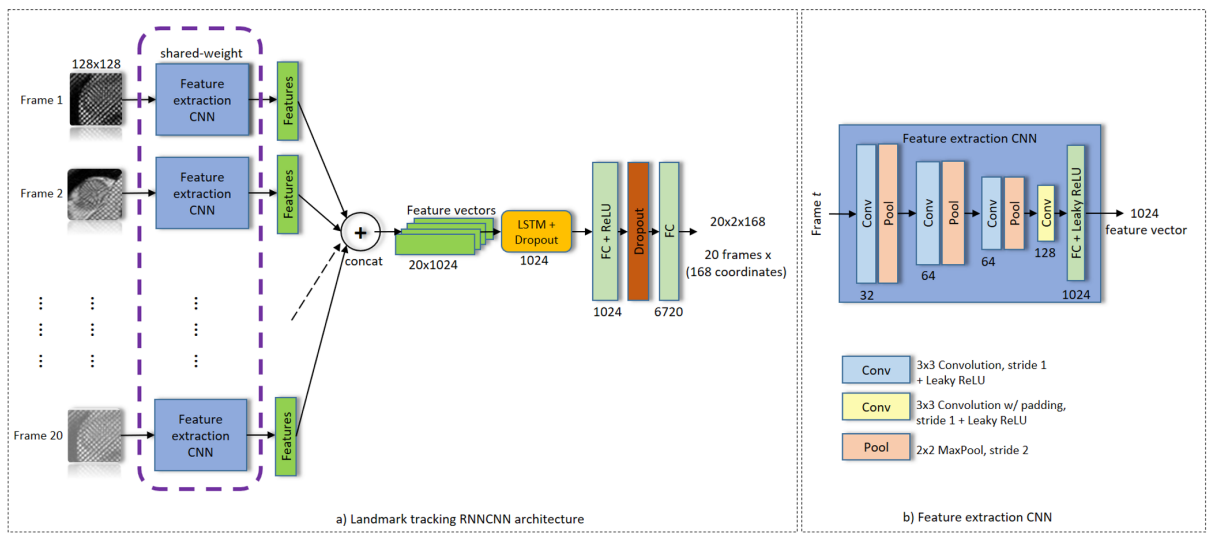

**Supplementary Figure 2** Landmark tracking RNNCNN (a) with a shared-weight spatial feature extraction component using CNN and temporal feature extraction using LSTM (Long-Short Term Memory), feature extraction component (b).